\documentclass[fleqn,usenatbib]{mnras}

\usepackage{newtxtext,newtxmath}

\usepackage[T1]{fontenc}

\DeclareRobustCommand{\VAN}[3]{#2}
\let\VANthebibliography\thebibliography
\def\thebibliography{\DeclareRobustCommand{\VAN}[3]{##3}\VANthebibliography}

\usepackage{graphicx}	
\usepackage{amsmath}	
\usepackage{amssymb}	

\newlength{\onecolumnwidth}
\usepackage{color}
\usepackage[normalem]{ulem}
\definecolor{darkgreen}{rgb}{0.13, 0.55, 0.13}

\usepackage{scalerel,tikz}
\usetikzlibrary{svg.path}
\definecolor{orcidlogocol}{HTML}{A6CE39}
\tikzset{orcidlogo/.pic={
 \fill[orcidlogocol] svg{M256,128c0,70.7-57.3,128-128,128C57.3,256,0,198.7,0,128C0,57.3,57.3,0,128,0C198.7,0,256,57.3,256,128z};
 \fill[white] svg{M86.3,186.2H70.9V79.1h15.4v48.4V186.2z}
 svg{M108.9,79.1h41.6c39.6,0,57,28.3,57,53.6c0,27.5-21.5,53.6-56.8,53.6h-41.8V79.1z M124.3,172.4h24.5c34.9,0,42.9-26.5,42.9-39.7c0-21.5-13.7-39.7-43.7-39.7h-23.7V172.4z}
 svg{M88.7,56.8c0,5.5-4.5,10.1-10.1,10.1c-5.6,0-10.1-4.6-10.1-10.1c0-5.6,4.5-10.1,10.1-10.1C84.2,46.7,88.7,51.3,88.7,56.8z};
}}
\newcommand\orcidicon[1]{\href{https://orcid.org/#1}{\mbox{\scalerel*{
\begin{tikzpicture}[yscale=-1,transform shape]
\pic{orcidlogo};
\end{tikzpicture}
}{|}}}}

\title[The local interstellar environment]{Formation and evolution of the local interstellar environment: combined constraints from nucleosynthetic and X-ray data}

\author[Y. Fujimoto et al.]{Yusuke Fujimoto$^{1}$\thanks{E-mail: yfujimoto@carnegiescience.edu}~\orcidicon{0000-0002-2107-1460},
Mark R. Krumholz$^{2,3}$~\orcidicon{0000-0003-3893-854X},
Shu-ichiro Inutsuka$^{4}$~\orcidicon{0000-0003-4366-6518},
\newauthor
Alan P. Boss$^{1}$~\orcidicon{0000-0001-7119-1105},
and Larry R. Nittler$^{1}$~\orcidicon{0000-0002-5292-6089}
\\
$^{1}$Earth and Planets Laboratory, Carnegie Institution for Science, 5241 Broad Branch Road, NW, Washington, DC 20015, USA\\
$^{2}$Research School of Astronomy and Astrophysics, Australian National University, Canberra 2611, A.C.T., Australia\\
$^{3}$ARC Centre of Excellence for Astronomy in Three Dimensions (ASTRO-3D), Canberra 2611, A.C.T., Australia\\
$^{4}$Department of Physics, Nagoya University, Furo-cho, Chikusa-ku, Nagoya, Aichi 464-8602, Japan
}

\date{Accepted XXX. Received YYY; in original form ZZZ}

\pubyear{2020}

\begin{document}
\label{firstpage}
\pagerange{\pageref{firstpage}--\pageref{lastpage}}
\maketitle

\begin{abstract}
Several observations suggest that the Solar system has been located in a region affected by massive stellar feedback for at least a few Myr; these include detection of live $^{60}\text{Fe}$ in deep-sea archives and Antarctic snow, the broad angular distribution of $^{26}{\rm Al}$ around the Galactic plane seen in all-sky $\gamma$-ray maps, and the all-sky soft X-ray background.
However, our position inside the Galactic disc makes it difficult to fully characterise this environment, and our limited time baseline provides no information about its formation history or relation to large-scale Galactic dynamics.
We explore these questions by using an $N$-body+hydrodynamics simulation of a Milky-Way-like galaxy to identify stars on Sun-like orbits whose environments would produce conditions consistent with those we observe. 
We find that such stars are uncommon but not exceptionally rare. These stars are found predominantly near the edges of spiral arms, and lie inside kpc-scale bubbles that are created by multiple generations of star formation in the arm.
We investigate the stars' trajectories and find that the duration of the stay in the bubble ranges from 20 Myr to 90 Myr.
The duration is governed by the crossing time of stars across the spiral arm. This is generally shorter than the bubble lifetime, which is $\sim 100$ Myr as a result of the continuous gas supply provided by the arm environment.
\end{abstract}

\begin{keywords}
Galaxy: general -- stars: massive -- ISM: bubbles -- Earth -- gamma-rays: ISM -- X-rays: ISM
\end{keywords}



\section{Introduction}

The Solar system is embedded in a low-density, warm, and partially ionised interstellar medium (ISM). The local ISM is dominated by the large cavity known as the Local Hot Bubble (LHB), whose existence was first suggested by extinction mapping, which showed  that stars within $\sim 100$ pc of the Sun experience negligible reddening (see the review by \citealt{Frisch_Redfield_Slavin_2011}, and references therein), implying very low local dust densities. Further evidence for a low-density cavity has come from all-sky soft X-ray surveys, which reveal a significant high-latitude background at $\sim 1/4$ keV  \citep[e.g.][]{Snowden_et_al_1995, Liu_et_al_2017}. The most likely explanation for this excess is the presence of a bubble of hot, low-density, ionised gas around or close to the Solar system, formed by stellar winds or supernovae (SNe) from nearby massive stars. 

Detections of live $^{60}\text{Fe}$ in deep-sea archives, lunar regolith,  and Antarctic snow provide a completely independent line of evidence that SNe occurred within $\sim$ 100 pc of the Earth within the last few Myr. $^{60}\text{Fe}$ is a radioactive element whose half-life is 2.6 Myr. It is synthesized in the late stages of massive stellar evolution and then ejected into the ISM by SNe \citep[e.g.][]{Lugaro_Ott_Kereszturi_2018}. The first discovery of $^{60}\text{Fe}$ was in ferro-manganese (Fe-Mn) crust from the South Pacific \citep{Knie_et_al_1999}, and a distinct signal in $^{60}\text{Fe}$ abundance a few Myr ago was confirmed by many other deep-sea archives of Fe-Mn crusts, sediments and nodules taken from all major oceans \citep{Knie_et_al_2004, Fitoussi_et_al_2008, Wallner_et_al_2016, Ludwig_et_al_2016}. Lunar regolith samples from the Apollo missions also show the presence of an excess of $^{60}\text{Fe}$ on the surface of the moon \citep{Fimiani_et_al_2016}. A direct detection of $^{60}\text{Fe}$ nuclei in Galactic cosmic rays also supports a recent near-Earth SN within a few Myr \citep{Binns_et_al_2016}. The Scorpius–Centaurus OB association and the Tucana-Horologium OB association have been suggested as a possible source of the $^{60}\text{Fe}$-producing SNe \citep{Breitschwerdt_et_al_2016, Schulreich_et_al_2017, Hyde_Pecaut_2018}. Moreover, a detection of $^{60}\text{Fe}$ in Antarctic snow shows that delivery of this isotope was not limited to a single event in the geologic past. Explaining the Antarctic snow requires a $^{60}\text{Fe}$ flux onto Earth over the last 20 yr, indicating that the Solar system is currently traversing an $^{60}\text{Fe}$-rich ISM contaminated by one or more SNe that occurred within the last few Myr \citep{Koll_et_al_2019, Koll_et_al_2020}.

A third line of evidence for the influence of nearby massive stellar feedback comes from the distribution of $^{26}\text{Al}$ seen in all-sky $\gamma$-ray maps. $^{26}\text{Al}$ is another radioactive element, with a half-life of 0.7 Myr, that is injected into the ISM either by SNe or by the stellar winds that precede the explosion \citep[e.g.][]{Lugaro_Ott_Kereszturi_2018}. Interstellar $^{26}\text{Al}$ has been observed with the 1809 keV $\gamma$-ray emission line, which traces downward nuclear transitions in the excited $^{26}\text{Mg}$ nuclei left behind when $^{26}\text{Al}$ decays. The Galactic sky-map of $^{26}\text{Al}$ shows a broad distribution; $^{26}\text{Al}$ extends to Galactic latitude $b > 5^{\circ}$ \citep{Pluschke_et_al_2001, Bouchet_Jourdain_Roques_2015}, while molecular gas \citep[e.g.][]{Dame_et_al_2001, Umemoto_et_al_2017} and young, massive stars \citep[e.g.][]{Bobylev_Bajkova_2016, Anderson_et_al_2019, Cantat-Gaudin_et_al_2020} are confined to $b < 2^{\circ}$. While some authors have suggested that the large angular scale height of $^{26}{\rm Al}$ reflects the true Galactic distribution \citep[e.g.][]{Wang_et_al_2020}, \citet{Fujimoto_Krumholz_Inutsuka_2020} show that a more plausible scenario is that the $^{26}{\rm Al}$ signal off the plane is the product of foreground emission from the $^{26}\text{Al}$ produced by a recent nearby SNe \citep[see also][]{Pleintinger_et_al_2019}.

These observations characterise the instantaneous properties of the local ISM, but tell us little about how such environments form or evolve over time, or how they are related to galactic-scale gas and stellar dynamics. Due to the long timescales involved, such questions can be addressed only through numerical simulations. Although \citet{Breitschwerdt_et_al_2016} and \citet{Schulreich_et_al_2017} succeeded in reproducing detailed structures of the LHB and individual events of $^{60}\text{Fe}$ transport from nearby SNe using hydrodynamical simulations of the local supperbubble, their simulations are purely local and are hand-tailored to reproduce the local environment, and thus cannot address the questions on which we focus here: how common are such environments like the Sun's within the Galactic disc? Where within the disc are they likely to be found? Once formed, how long do such environments persist? In this paper, we attempt to answer these Galactic-scale statistical questions using the combination of astrophysical soft X-ray and $^{26}{\rm Al}$ data, and terrestrial $^{60}{\rm Fe}$ data. To the end, we use the $N$-body+hydrodynamics simulation of a Milky-Way-like galaxy model of \citet{Fujimoto_Krumholz_Inutsuka_2020}. This model has enough resolution and physics to track the bubbles of hot gas and short-lived radioisotopes produced by individual supernovae, and includes a live disc of older stars such as the Sun. The simulation therefore allows us to construct realistic estimates of the distribution of $^{60}{\rm Fe}$, $^{26}{\rm Al}$, and X-ray sky background that would be seen from a large sample of Sun-like stars. We use this capability to investigate the formation and evolution of the environments of stars whose properties are consistent with the observational constraints we have for the present-day environment of the Sun.

This paper is organised as follows. In Section~\ref{Methods}, we briefly summarise our numerical model of a Milky-Way-like galaxy. In Section~\ref{Results}, first we select Sun-like motion stars that meet the three observational constraints: $^{60}{\rm Fe}$ flux onto the Earth, $^{26}{\rm Al}$ scale latitude observed in $\gamma$-ray sky-maps, and the mean flux of diffuse soft X-ray emission. Next we discuss the location of the stars in the Galactic disc and the time evolution of the local interstellar environment. In Section~\ref{Conclusions and Discussion}, we summarise our findings.

\section{Methods}
\label{Methods}

We carry out this project using the $N$-body+hydrodynamics simulation of a Milky-Way-like galaxy described in \citet{Fujimoto_Krumholz_Inutsuka_2020}. We refer readers to that paper for full details of the numerical method, and here simply summarise the most important aspects of the simulation. In this paper, we use the result at $t = $ 650 Myr.

The dark matter and stars are represented by collisionless particles in the galaxy model with $N_{\text{halo}} = 10^7$ dark matter halo particles, $N_{\text{disc}} = 10^7$ stellar disc particles, and $N_{\text{bulge}} = 1.25 \times 10^6$ stellar bulge particles. All particles in each population have uniform masses: $m_{\mathrm{halo}} = 1.254 \times 10^5\ \mathrm{M_{\sun}}$ for the halo population and $m_{\mathrm{disc}} = m_{\mathrm{bulge}} = 3.437 \times 10^3\ \mathrm{M_{\sun}}$ for the disc and bulge populations. The initial gas distributions on the grid structure are initialized following an analytic exponential density profile. The mass distribution of all the four components (halo, disc, bulge and gas) sets an initial rotation curve of the gas disc with circular velocity $V_{\mathrm{c, gas}}(R = 8\ \mathrm{kpc}) = 237\ \mathrm{km\ s^{-1}}$, consistent with observations \citep[e.g.][]{Reid_et_al_2019}. We set the initial abundances of $^{60}\text{Fe}$ and $^{26}\text{Al}$ to $10^{-12}$ throughout the simulation box, though this choice has no practical effect since the initial abundances decay rapidly.

Our simulation follows the evolution of $N$-body particles and hydrodynamic gas using the adaptive mesh refinement (AMR) code \textsc{Enzo} \citep{Bryan_et_al_2014}. We treat $^{60}\text{Fe}$ and $^{26}\text{Al}$ as passive scalars that are transported with the gas, and that decay with half-lives of $t_{1/2} = 2.62$ Myr for $^{60}\text{Fe}$ and $t_{1/2} = 0.72$ Myr for $^{26}\text{Al}$ \citep{Rugel_et_al_2009, Norris_et_al_1983}. Here we assume that the $^{60}\text{Fe}$ and $^{26}\text{Al}$ dust grains (e.g. Fe$_3$O$_4$ and Al$_2$O$_3$) and gas are well coupled at the spatial scale we resolve in this simulation because the typical drift velocity of the small dust ($\leq$ 10 $\mu$m) relative to gas at a hundred parsec scale in the galactic disc is much smaller than the typical turbulent velocity of the ISM (see Appendix~\ref{Dust-gas coupling}), and the typical sizes of Al and Fe dust grains in SN ejecta are predicted to be less than 0.01 $\mu$m by theoretical models \citep{Bocchio_et_al_2016}. Meteoritic measurements indicate that grains produced by SNe can range in size up to tens of microns \citep[e.g.][]{Gyngard_et_al_2018}, but most are probably of order 100 nm or smaller \citep{Hoppe_Leitner_Kodolanyi_2015}. Decoupling of grains and gas, and interactions between charged grains and the Solar magnetic field, might affect the eventual influx of materials onto the Earth \citep{Fry_et_al_2020}. We account for this effect approximately when we compute the rate of $^{60}{\rm Fe}$ deposition on Earth, but do not otherwise attempt to include it, since we do not resolve scales small enough for it to be important.

The galaxy is modelled in a 3D simulation box of $(1.31072\ \mathrm{Mpc})^3$. The root grid is $64^3$ cells, on top of which we impose another 5 levels of statically refined regions. As a result, the galactic disc is enclosed within a $(40.96\ \mathrm{kpc})^3$ box with cells 640 pc in size. In addition to the static refinement, we impose an additional 5 levels of adaptive refinement, producing a minimum cell size of $\Delta x = 20$ pc.

We include stochastic star formation and stellar feedback from photoionization, SNe and stellar winds, as well as chemical injections of $^{60}\text{Fe}$ and $^{26}\text{Al}$. Star particles form in gas where the number density exceeds 13 $\mathrm{cm}^{-3}$, corresponding to the density for which gas at the equilibrium temperature set by our cooling curve becomes Jeans unstable at our peak resolution of 20 pc. The star formation efficiency per free-fall time in star-forming gas is 0.01. Rather than spawning star particles in every cell at each time-step, we form particles stochastically imposing a minimum star particle mass of 300 $\text{M}_{\sun}$. To model SN, wind, and photoionisation feedback from massive stars, we use the \textsc{slug} stellar population synthesis code \citep{Krumholz_et_al_2015}; each star particle spawns an individual \textsc{slug} simulation that stochastically draws individual stars from the initial mass function (IMF), tracks their mass- and age- dependent ionising luminosities and stellar wind mechanical luminosities, determines when individual stars explode as SNe, and calculates the resulting injection of $^{60}\text{Fe}$ and $^{26}\text{Al}$. In the \textsc{slug} calculation, we use a Chabrier IMF \citep{Chabrier_2005} with \textsc{slug}'s Poisson sampling option, Padova stellar evolution tracks with Solar metallicity \citep{Girardi_et_al_2000}, \textsc{starburst99} stellar atmospheres \citep{Leitherer_et_al_1999}, and the mass-dependent yield table of \citet{Sukhbold_et_al_2016}. Because of the limited computational resources and time, we do not explore alternative models of stellar evolution and SN yields. However, it would be worthwhile repeating our simulations in the future with other models and parameters, in particular alternative models of chemical yields of $^{60}\text{Fe}$ and $^{26}\text{Al}$ \citep[e.g.][]{Chieffi_Limongi_2013}, to investigate the model dependence.

\section{Results}
\label{Results}

\subsection{Selection of stars that meet observational constraints}

\begin{figure*}
    \centering
	\includegraphics[width=0.33\textwidth]{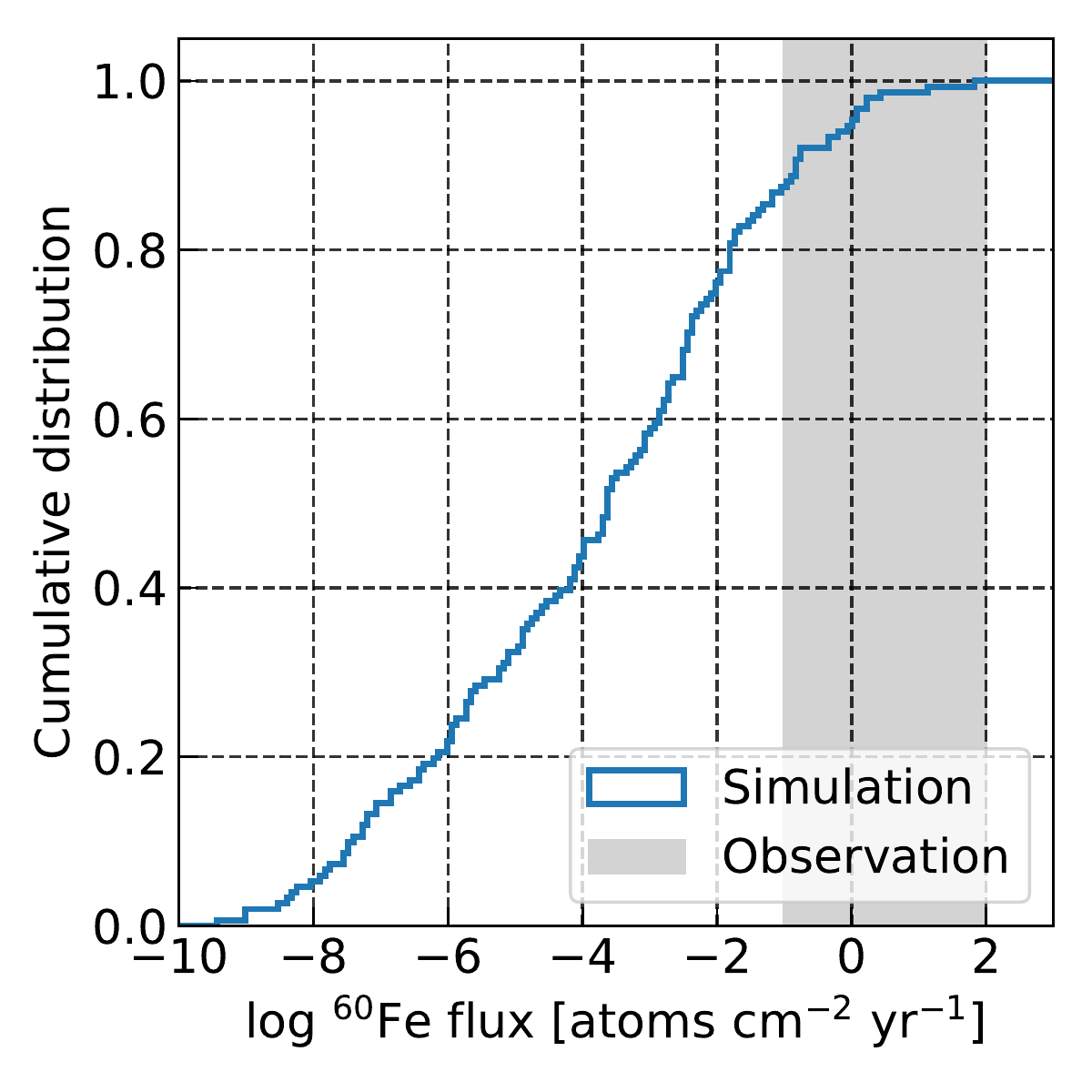}
	\includegraphics[width=0.33\textwidth]{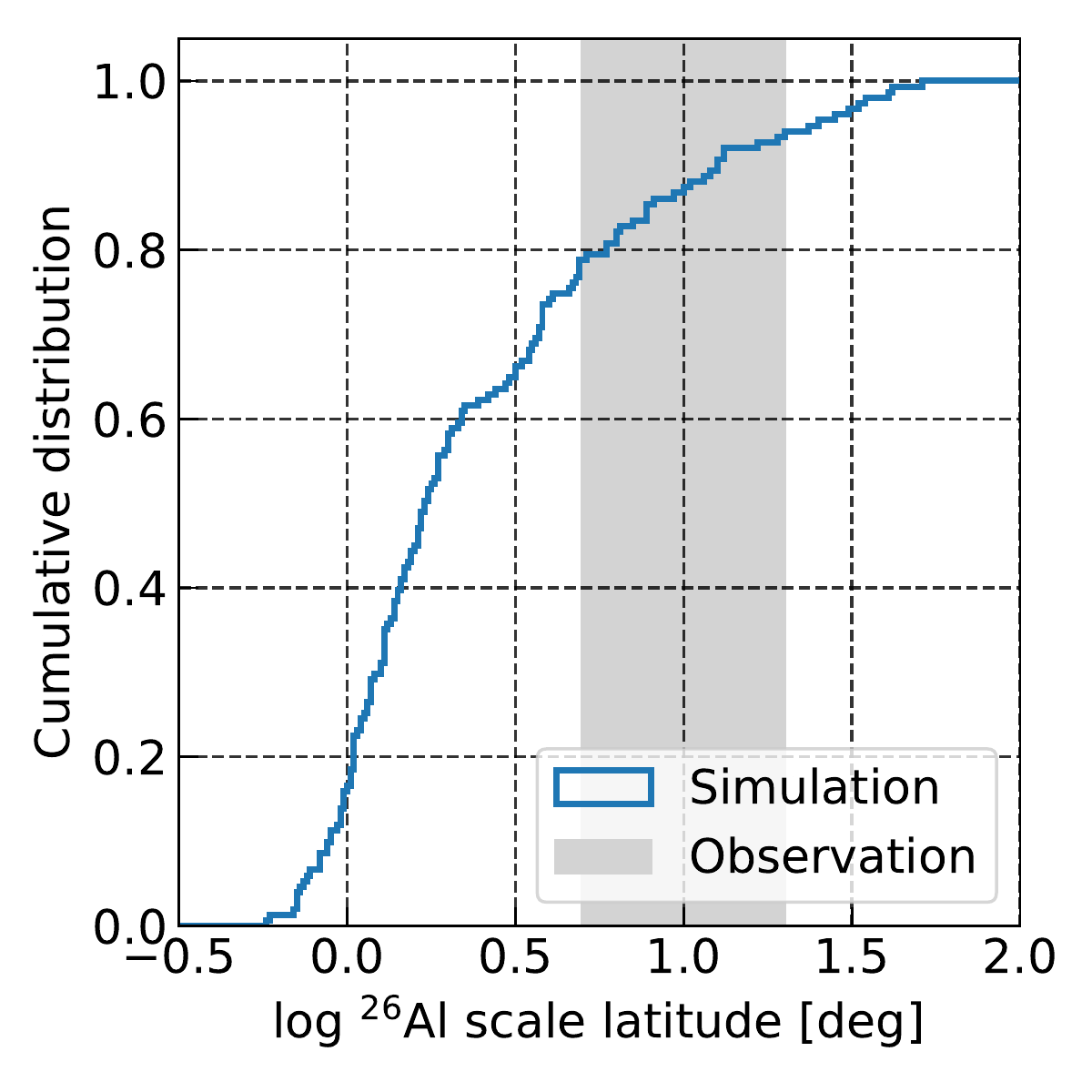}
	\includegraphics[width=0.33\textwidth]{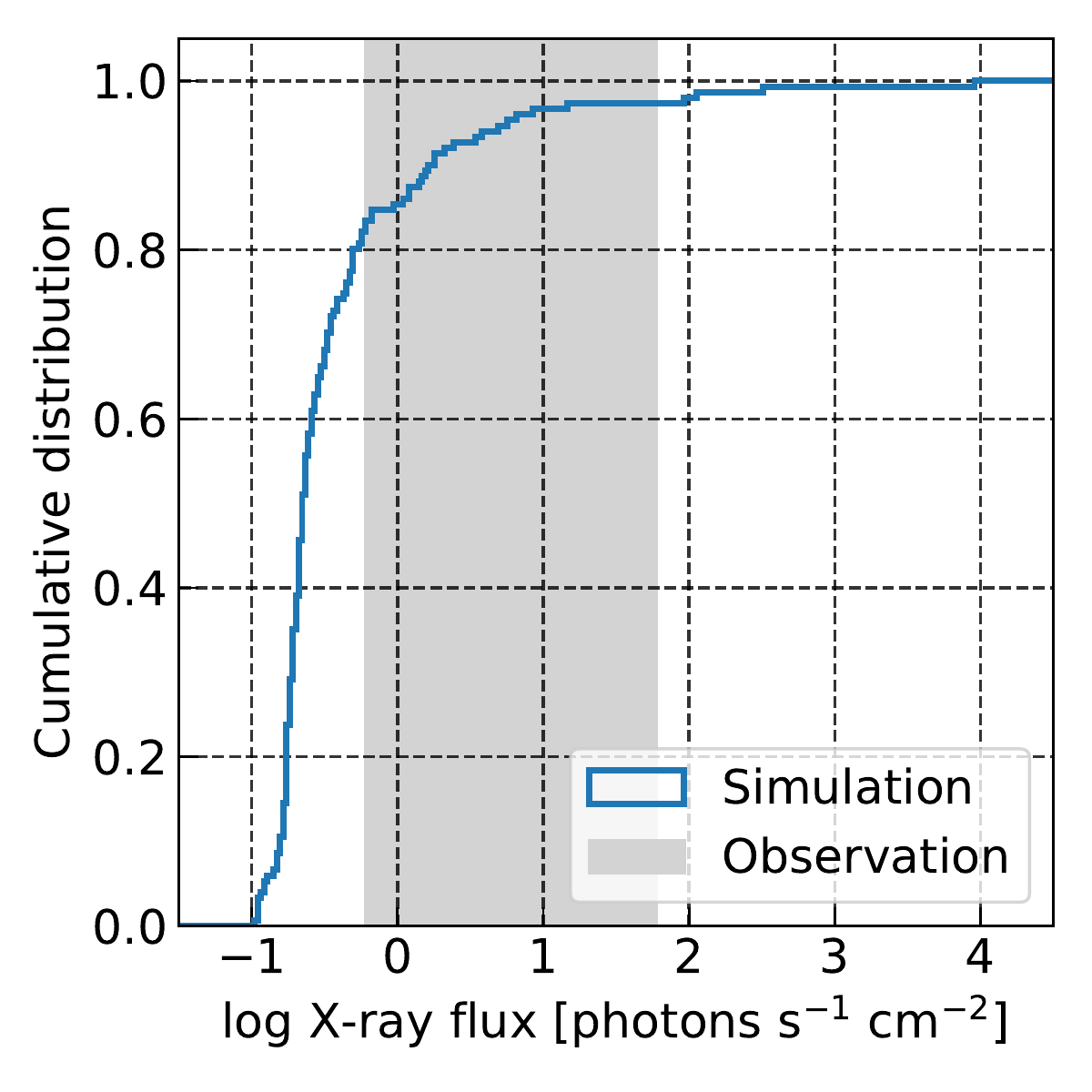}
    \caption{Normalised cumulative distribution functions of $^{60}\text{Fe}$ flux (left), $^{26}\text{Al}$ scale latitude (centre), and X-ray flux (right), for sample stars in the simulation. The observed ranges are shown in grey.
    }
    \label{fig: PDFs}
\end{figure*}

\begin{figure*}
    \centering
	\includegraphics[width=0.33\textwidth]{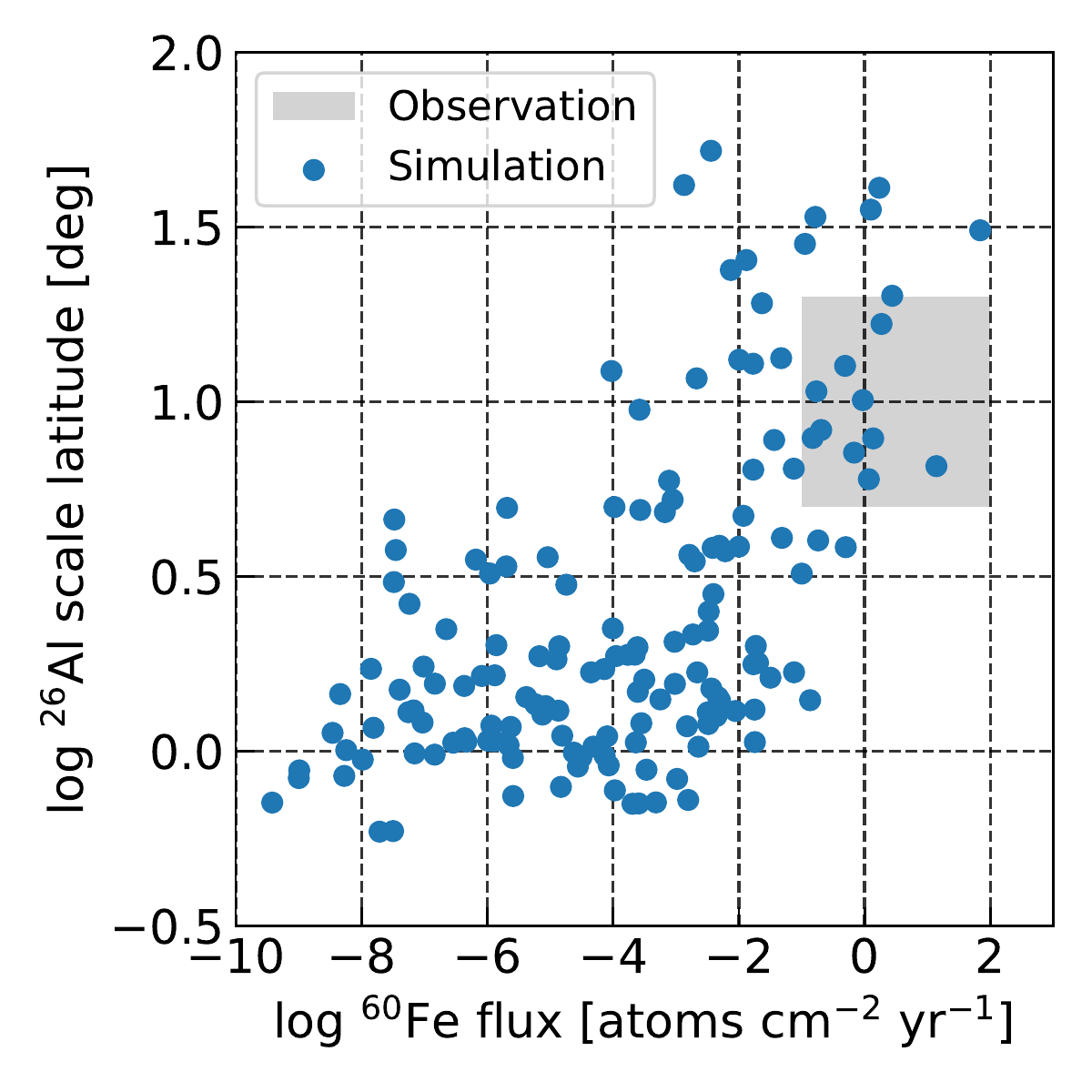}
	\includegraphics[width=0.33\textwidth]{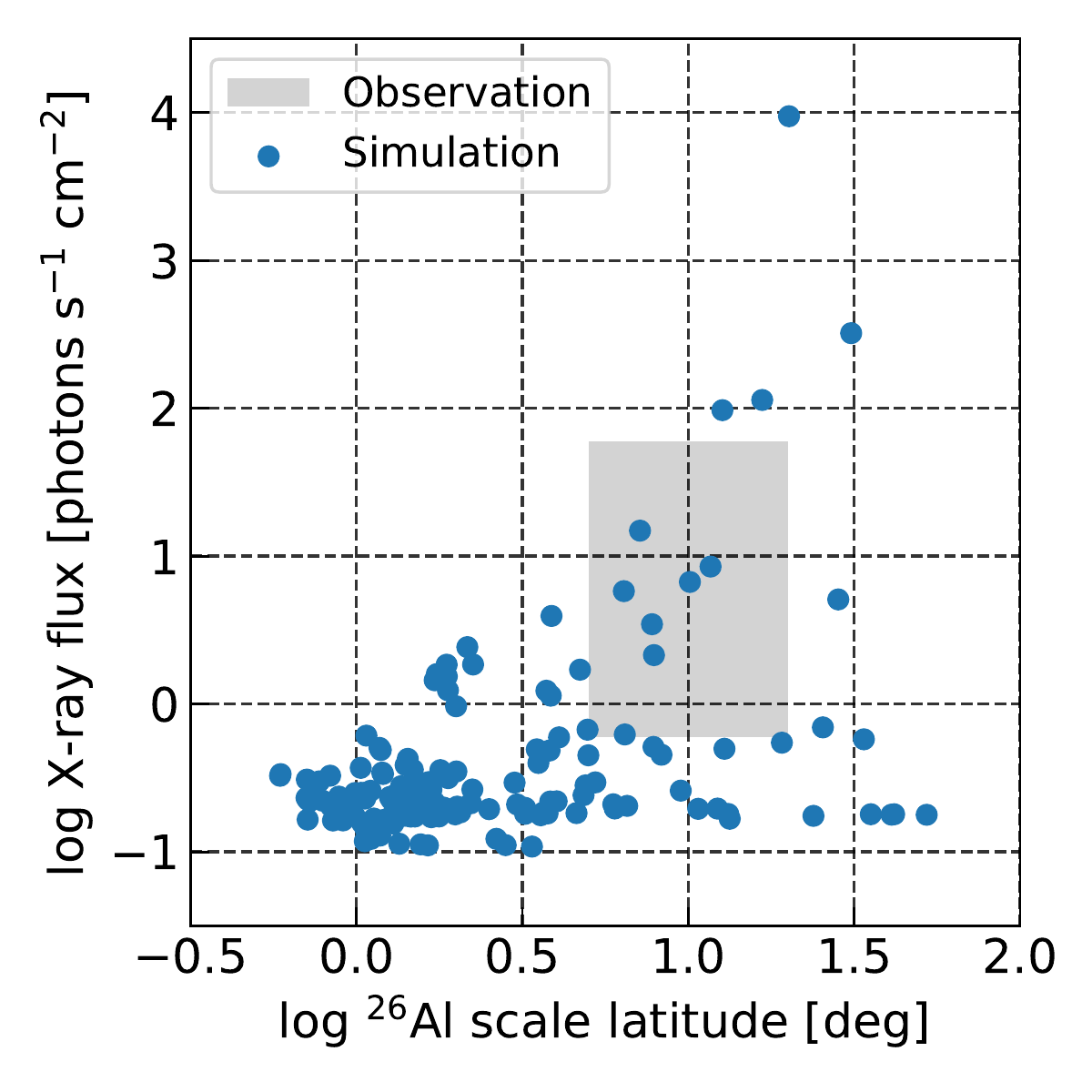}
	\includegraphics[width=0.33\textwidth]{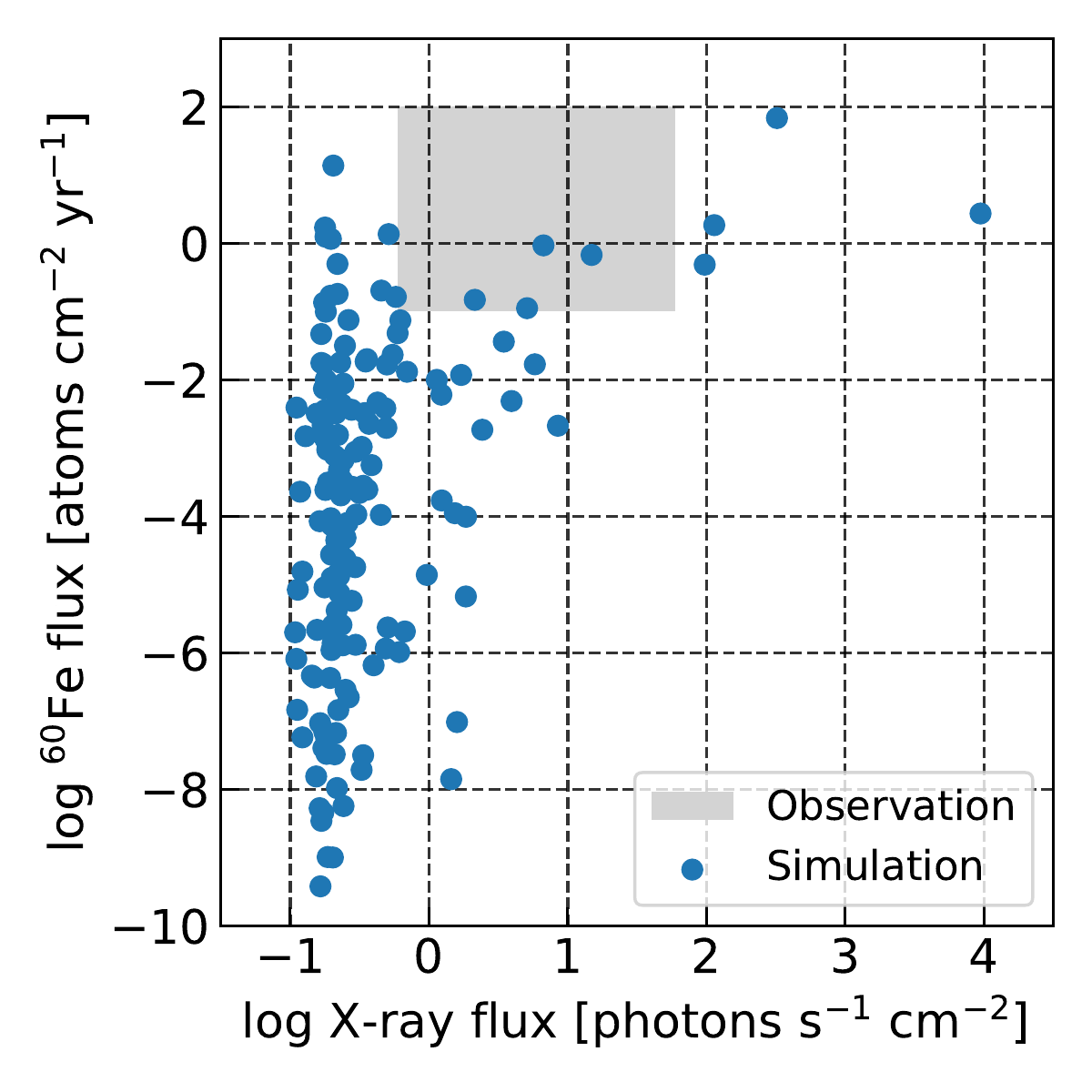}
    \caption{Scatter plots for $^{60}\text{Fe}$ flux vs. $^{26}\text{Al}$ scale latitude (left), $^{26}\text{Al}$ scale latitude vs. X-ray flux (centre), and X-ray flux vs. $^{60}\text{Fe}$ flux (right), for sample stars in the simulation. The observational ranges are shown in grey. 
    }
    \label{fig: scaling_relations}
\end{figure*}

Our first step in analysing the simulation is to select stars on Sun-like orbits, since we might reasonably expect the distribution of $^{60}{\rm Fe}$, $^{26}{\rm Al}$, and X-ray emitting hot gas around stars to depend on their position in the Galactic disc. We select our sample based on three criteria: Galactocentric distances between 8.1 kpc $< R <$ 8.3 kpc, offsets from the mid-plane of the Galactic disc between 20 pc $< |z| <$ 30 pc, and vertical velocities perpendicular to the Galactic disc between 6.5 km s$^{-1}$ $< |v_z| <$ 7.5 km s$^{-1}$ \citep{Bland-Hawthorn_Gerhard_2016}. The number of particles that meet these criteria is 151. In the following sections, we investigate which of these stars have observational indicators of their local interstellar environment consistent with those of the Sun.

\subsubsection{Constraint 1: $^{60}\text{Fe}$ flux onto Earth}

The first constraint to which we compare is the $^{60}\text{Fe}$ flux into the Solar system as recorded in deep-sea archives \citep{Knie_et_al_1999, Knie_et_al_2004, Fitoussi_et_al_2008, Wallner_et_al_2016, Ludwig_et_al_2016} and Antarctic snow \citep{Koll_et_al_2019} on Earth's surface, and in lunar regolith on Moon's surface \citep{Fimiani_et_al_2016}. Table II of \citet{Koll_et_al_2019} summarizes the $^{60}\text{Fe}$ fluxes inferred from a variety of samples; the data show a range of values roughly between $10^{-1}$ and $10^2$ atoms cm$^{-2}$ yr$^{-1}$.

To compare with the observations, we calculate the $^{60}\text{Fe}$ flux for each sample star in the simulation. We define the $^{60}\text{Fe}$ flux $\mathcal{F}$, as
\begin{equation}
    \mathcal{F} = \frac{f}{4} \frac{\rho_{^{60}\text{Fe}} |\mathbf{v}_{^{60}\text{Fe}}-\mathbf{v}_{\mathrm{star}}|}{Am_{\mathrm{u}}},
\end{equation}
where $\rho_{^{60}\text{Fe}}$ is the $^{60}\text{Fe}$ density in the cell that hosts the star, $\mathbf{v}_{^{60}\text{Fe}}$ is the velocity of the cell, $\mathbf{v}_{\mathrm{star}}$ is the star particle's velocity, $A = 60$ is the mass number of $^{60}\text{Fe}$, and $m_{\mathrm{u}}$ is the atomic mass unit. This is an analogous form to the equation 14 and 15 in \citet{Schulreich_et_al_2017} \citep[see also][]{Schulreich_2015}. The factor of 1/4 comes from relating Earth's cross section $(\upi R_{\earth}^2)$ to its surface area $(4 \upi R_{\earth}^2)$. The survival fraction $f$ is the fraction of $^{60}\text{Fe}$ atoms that are condensed into solid grains that reach the Earth's surface after overcoming a variety of filtering processes due to the magnetic, gravitational and radiative influences of the Sun and the Earth. We adopt a value of $f\simeq 0.01$ from \citet[also see \citealt{Fry_Fields_Ellis_2016}]{Fry_Fields_Ellis_2015}, who consider dust condensation at departure from source, dust destruction through SN remnants, and the filtering effects of passage through the heliosphere and solar radiation pressure.\footnote{Note that our flux $\mathcal{F}$ is referred to as the surface fluence or global mean fluence in \citet{Fry_Fields_Ellis_2015}, and that our $\mathcal{F}$ does not include a factor $U$ for the uptake efficiency of the atoms into some particular material. Thus our $\mathcal{F}$ represents the rate at which atoms arrive on the Earth's surface, without regard to their subsequent fate.} 

The left panel of \autoref{fig: PDFs} shows the cumulative distribution function (CDF) of the $^{60}\text{Fe}$ flux for our 151 Sun-like sample stars. We see that stars that have $^{60}{\rm Fe}$ fluxes within the observed range are located at the high-end tail of the CDF. Quantitatively, 19 of 151 stars (12.6 per cent) have $^{60}{\rm Fe}$ fluxes in the range $10^{-1}-10^2$ atoms cm$^{-2}$ s$^{-1}$ favoured by the observations.

\subsubsection{Constraint 2: $^{26}\text{Al}$ scale latitude}

The second constraint is a broad distribution of $^{26}\text{Al}$ extended to a high latitude of $5^{\circ} < b < 20^{\circ}$ observed in the $\gamma$-ray sky-map of $^{26}\text{Al}$ \citep{Pluschke_et_al_2001, Bouchet_Jourdain_Roques_2015}. To compare with the observation, we construct a synthetic $^{26}\text{Al}$ emission map of the sky as it would be seen from each sample star, and we calculate the scale latitude of $^{26}\text{Al}$,
\begin{equation}
    b_0 = \frac{\int^{90^{\circ}}_{0^{\circ}} \int^{180^{\circ}}_{-180^{\circ}} |b| \frac{\mathrm{d}M}{\mathrm{d}\Omega}(\ell,|b|) \mathrm{d}\ell \mathrm{d}|b|}{\int^{90^{\circ}}_{0^{\circ}} \int^{180^{\circ}}_{-180^{\circ}} \frac{\mathrm{d}M}{\mathrm{d}\Omega}(\ell,|b|) \mathrm{d}\ell \mathrm{d}|b|},
\end{equation}
where $\ell$ and $b$ are the longitude and latitude in Galactic coordinates as seen from a particular star, and $\mathrm{d}M/\mathrm{d}\Omega(\ell, |b|)$ is the $^{26}{\rm Al}$ mass per unit solid angle on the sky as viewed from the star.

The central panel of \autoref{fig: PDFs} shows the CDF of the $^{26}\text{Al}$ scale latitude. As seen in the case of $^{60}\text{Fe}$ flux, stars with scale latitudes as large as the one we observe on Earth are located at the high-end tail of the CDF, though not quite as far to the high end as was the case for the $^{60}{\rm Fe}$ flux. Quantitatively, 22 out of our 151 sample stars (14.6 percent) have $5^{\circ} < b < 20^{\circ}$.

\subsubsection{Constraint 3: soft X-ray flux}

Our third constraint is the mean flux of diffuse soft X-ray emission (1/4 keV) averaged over the whole sky. We compute this flux from the published ROSAT/XRT all-sky map of \citet{Snowden_et_al_1995}, masking out the region $-30^{\circ} < b < 30^{\circ}$ because regions near the midplane may be affected by absorption. Considering the 84-cm diameter aperture of ROSAT/XRT and assuming that 40 per cent of the observed emission arises from solar-wind charge-exchange rather than from the ISM \citep{Galeazzi_et_al_2014}, we arrive at a total mean flux of 6 photons s$^{-1}$ cm$^{-2}$ from the ISM. We consider a simulated star to be roughly consistent with this level of X-ray background if it has a sky-averaged flux in the range 0.6 - 60 photons s$^{-1}$ cm$^{-2}$, i.e., within a factor of 10 of the flux seen from Earth.

In order to determine which of our simulated stars would experience this level of X-ray background, we construct a synthetic X-ray emission map for each sample star. We generate these maps by assigning an emissivity to each cell based on its density and temperature, using tabulated emissivities computed from \textsc{Cloudy} \citep{Ferland_et_al_2013}, as implemented in \textsc{yt} \citep{Turk11a}. We then integrate the emission over angle to produce X-ray sky maps, and derive the mean flux from these maps using the same procedure that we apply to the observed map. 

The right panel of \autoref{fig: PDFs} shows the CDF of the X-ray flux. Again, stars that fall within our target X-ray background flux range are located at the high-end tail of the CDF. The number is 23 out of 151 total sample stars, and corresponding to a fraction of 15.2 per cent.

\subsubsection{Correlations among constraints}
\label{Correlations among constraints}

We have shown that stars that match the background of $^{60}{\rm Fe}$, $^{26}{\rm Al}$, and soft X-ray emission seen from Earth are not typical, but instead lie in the top $\approx 15\%$ of the CDFs for stars in Sun-like orbits. The next question to consider is how these observational constraints relate to each other; if they are uncorrelated, the odds for any given star to meet all three conditions would be only $12.6\% \times 14.6\% \times 15.2\% = 0.28\%$, and we would expect to find none in our sample of 151 stars in Sun-like orbits. \autoref{fig: scaling_relations} shows scatter plots among $^{60}\text{Fe}$ flux, $^{26}\text{Al}$ scale latitude, and X-ray flux, and demonstrates that this is not the case. Even though the three constraints come from completely different observations, there is a clear correlation between them. As a result we find 3 stars that meet all three conditions, corresponding to 2 per cent of the sample, which is one order of magnitude larger than the 0.28 per cent we would expect if the three constraints were uncorrelated.

\subsection{The local interstellar environment}

\begin{figure*}
    \centering
	\includegraphics[width=0.49\textwidth]{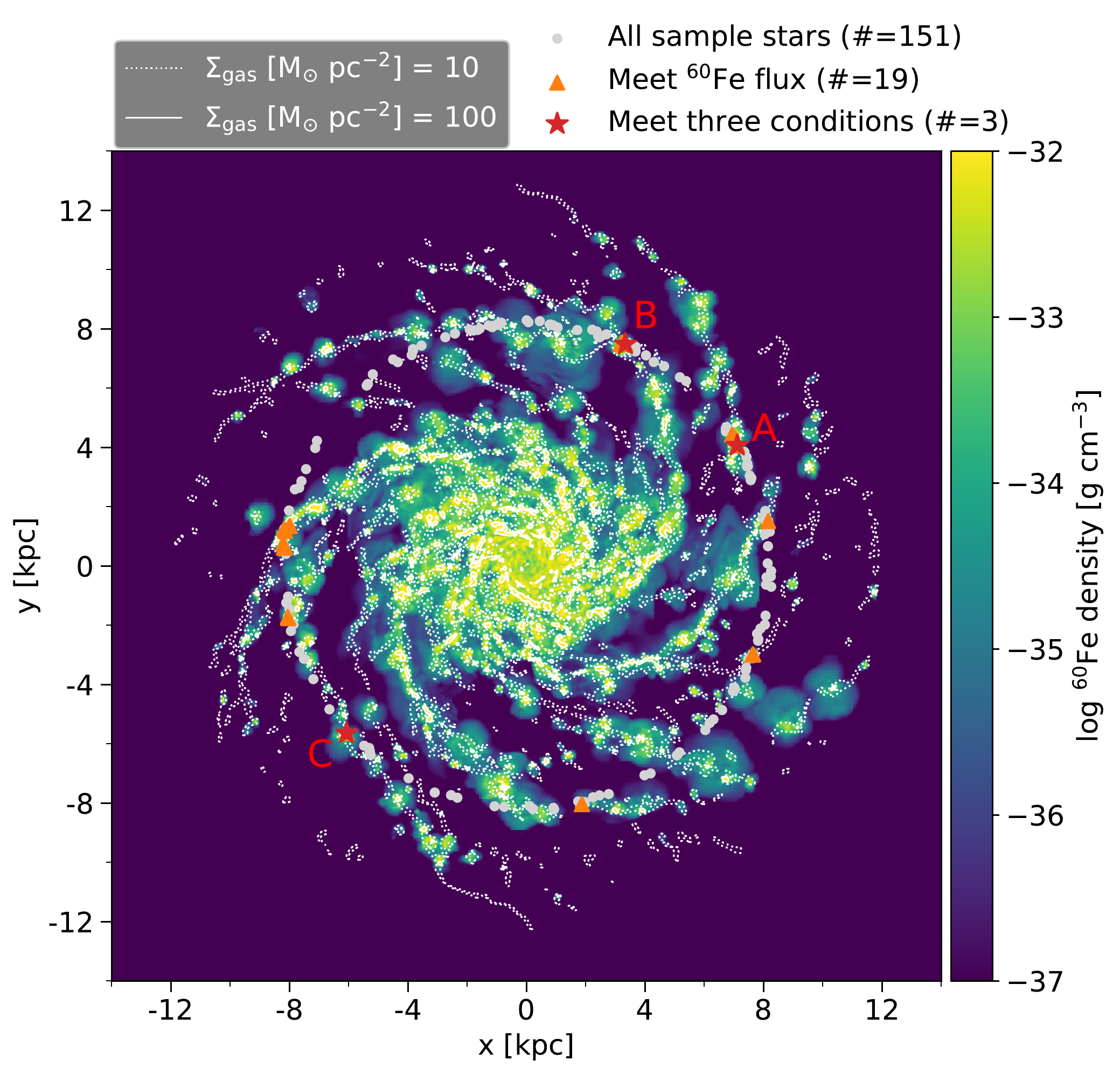}
	\includegraphics[width=0.49\textwidth]{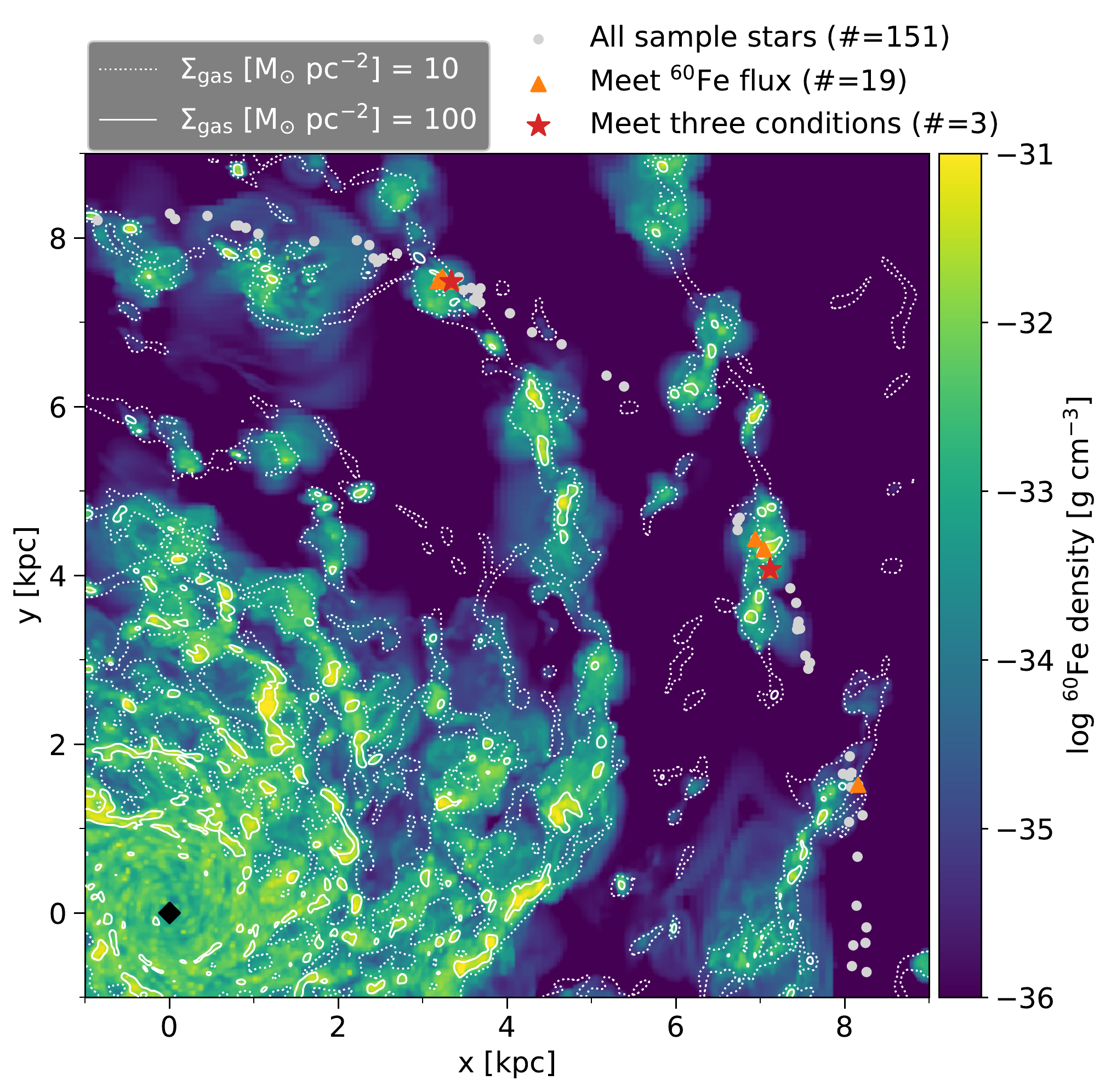}
	\includegraphics[width=0.49\textwidth]{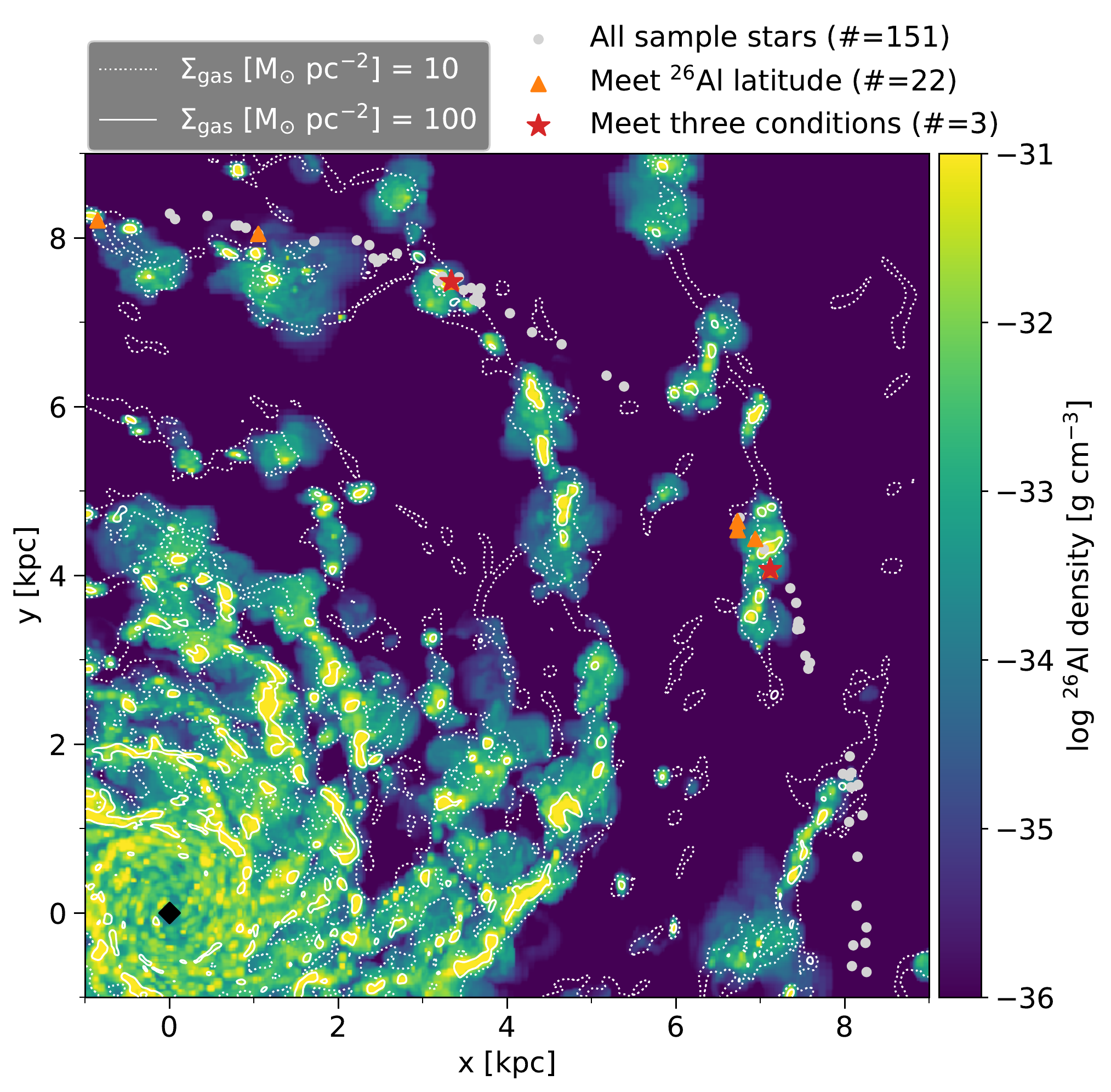}
	\includegraphics[width=0.49\textwidth]{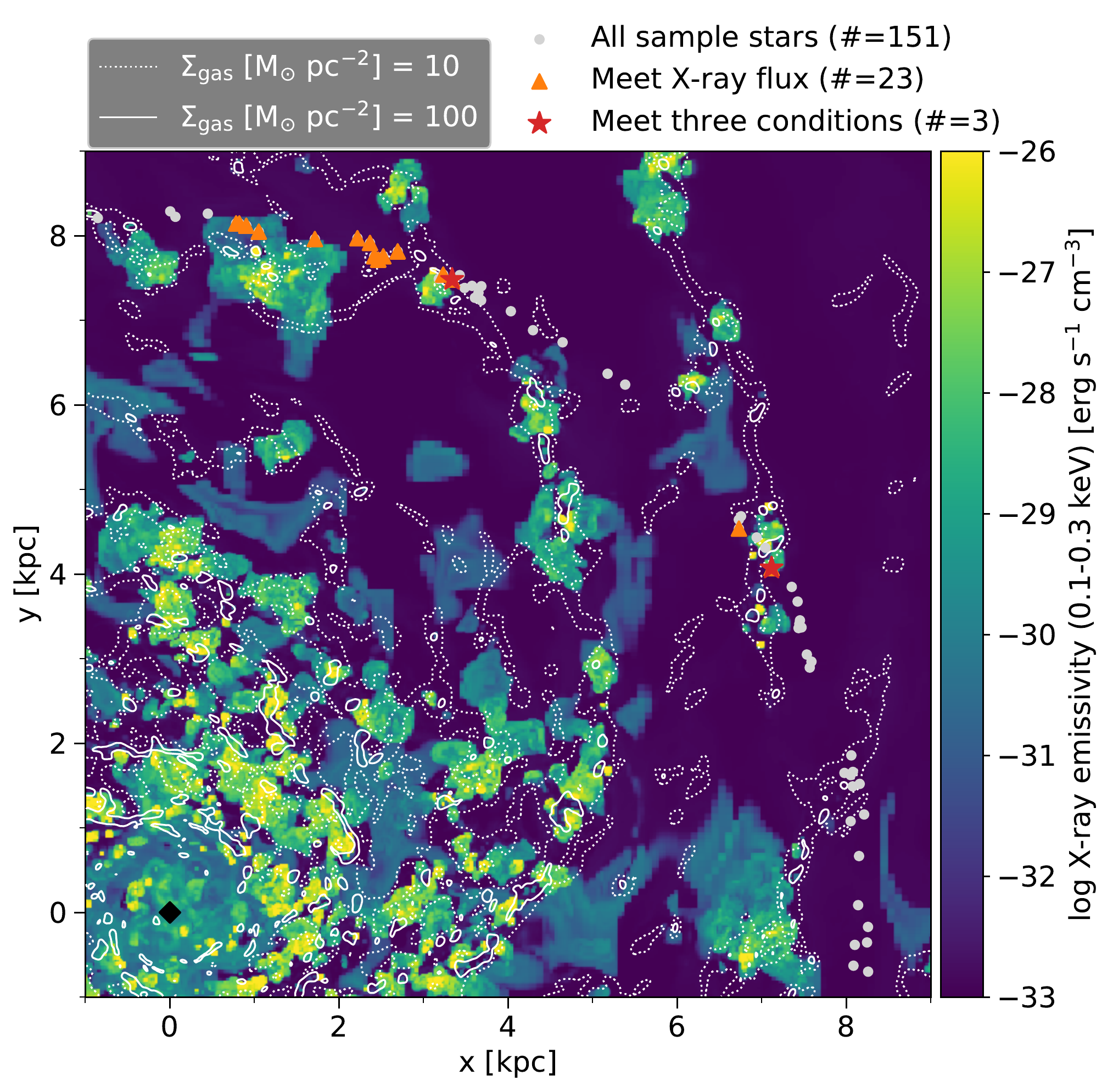}
    \caption{A face-on galactic disc image of $^{60}\text{Fe}$ (top left) and zoom-in images of $^{60}\text{Fe}$ (top right), $^{26}\text{Al}$ (bottom left) mass-weighted average densities, and X-ray emissivity (bottom right), integrated over $-250 < z < 250$ pc at $t = 650$ Myr, overlaid with the gas surface density (contours). The grey dots show our entire sample of stars on Sun-like orbits (8.1 kpc $< R <$ 8.3 kpc, 20 pc $< |z| <$ 30 pc, and 6.5 km s$^{-1}$ $< |v_z| <$ 7.5 km s$^{-1}$). The orange triangles show stars that satisfy one of the three constraints, and the red star marks show stars that meet all three conditions. The black diamond shows the galactic centre, and the galaxy rotates clockwise.}
    \label{fig: projections}
\end{figure*}

\begin{figure*}
    \centering
	\includegraphics[width=0.99\textwidth]{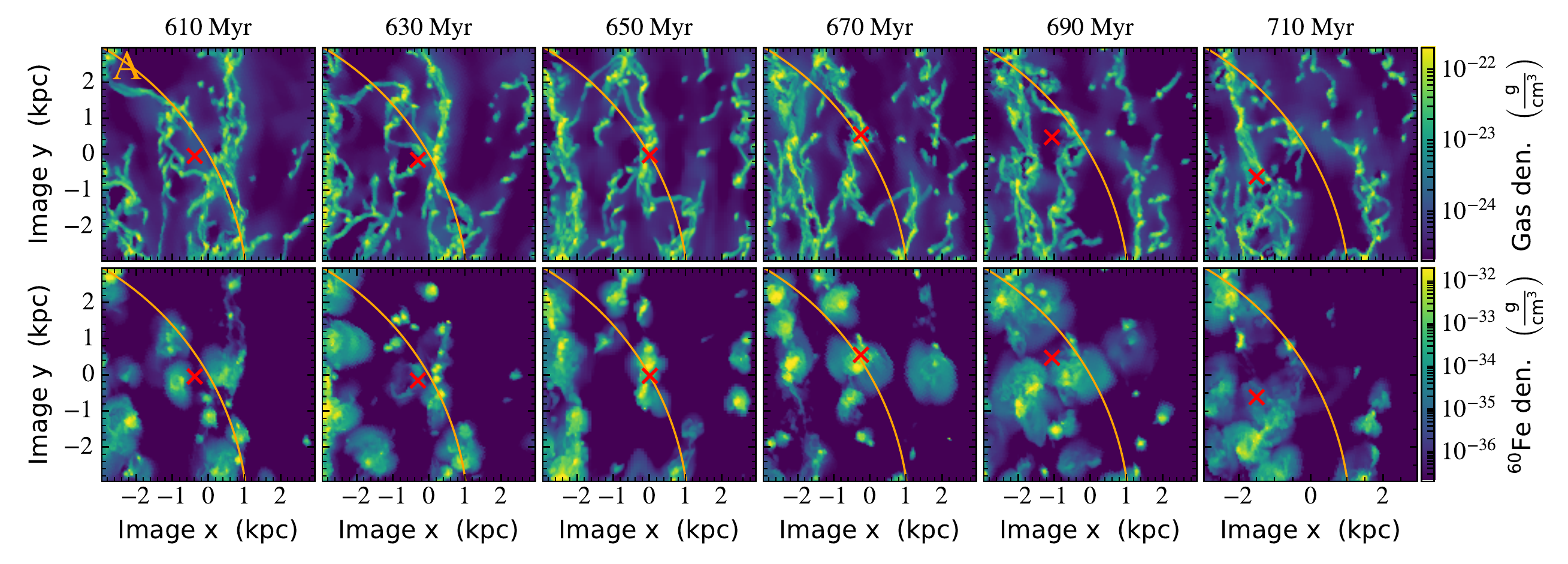}
	\includegraphics[width=0.99\textwidth]{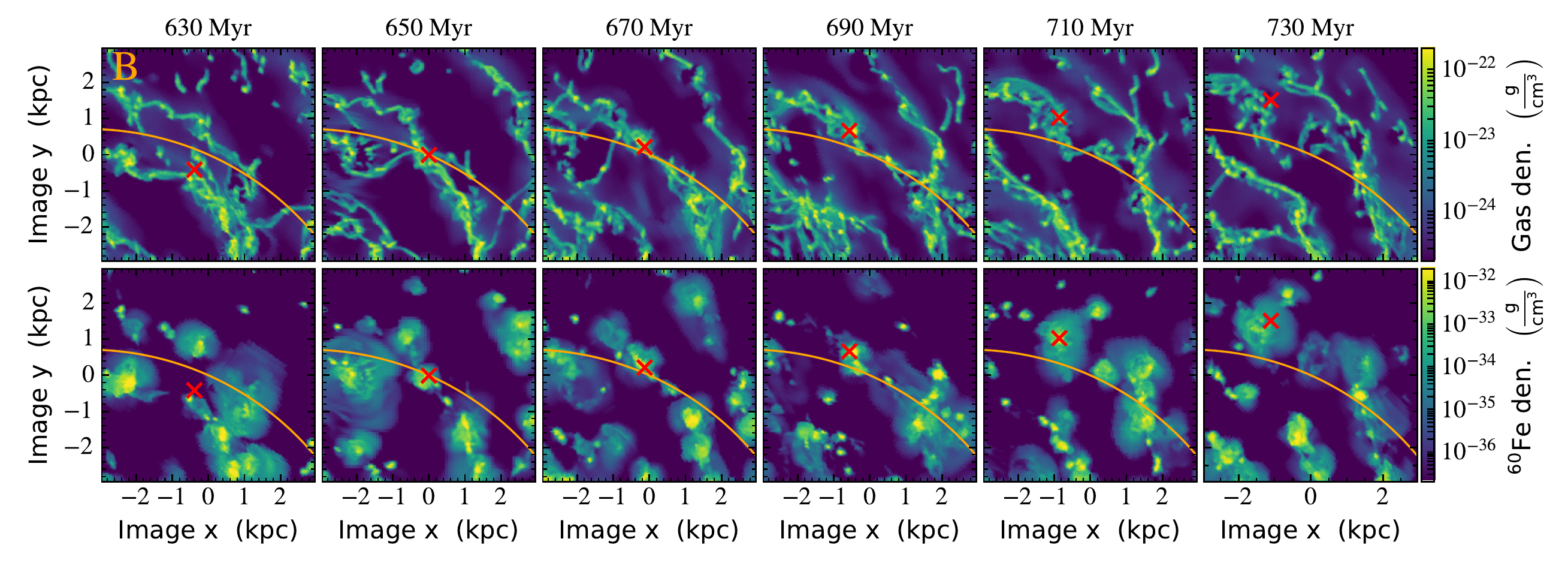}
	\includegraphics[width=0.99\textwidth]{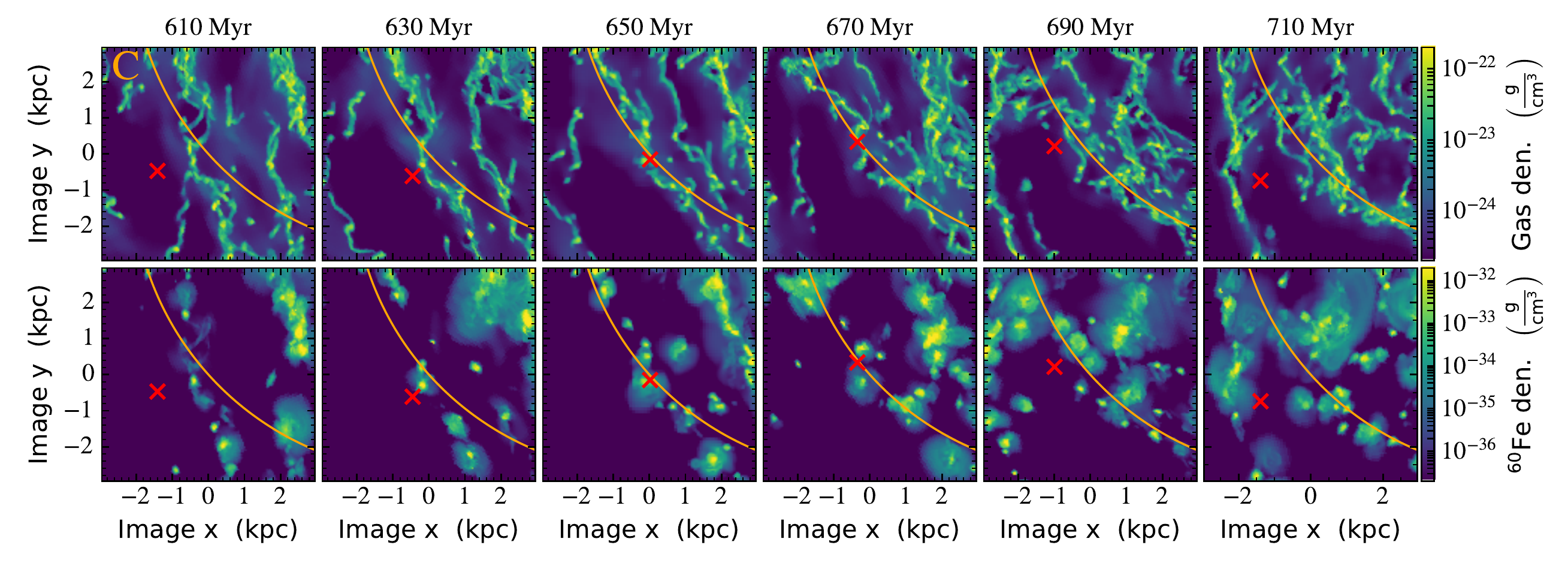}
    \caption{Time evolution of stars and their local interstellar environments that meet all three conditions. Panels of A (top), B (middle), and C (bottom) correspond to the three stars labelled A, B, and C in the top left panel of \autoref{fig: projections}. The top rows show the mass weighted gas density, and the bottom rows show the mass weighted $^{60}\text{Fe}$ density. The red marks show the positions of the stars. The orange arcs indicate a Galactocentric radius of 8.2 kpc and the figures are plotted in a frame that co-rotates with the Galactic rotation curve at $r = 8.2$ kpc. Movies are available with the online version of the journal.}
    \label{fig: time_evolution_bubbles}
\end{figure*}

\begin{figure}
    \centering
	\includegraphics[width=0.47\textwidth]{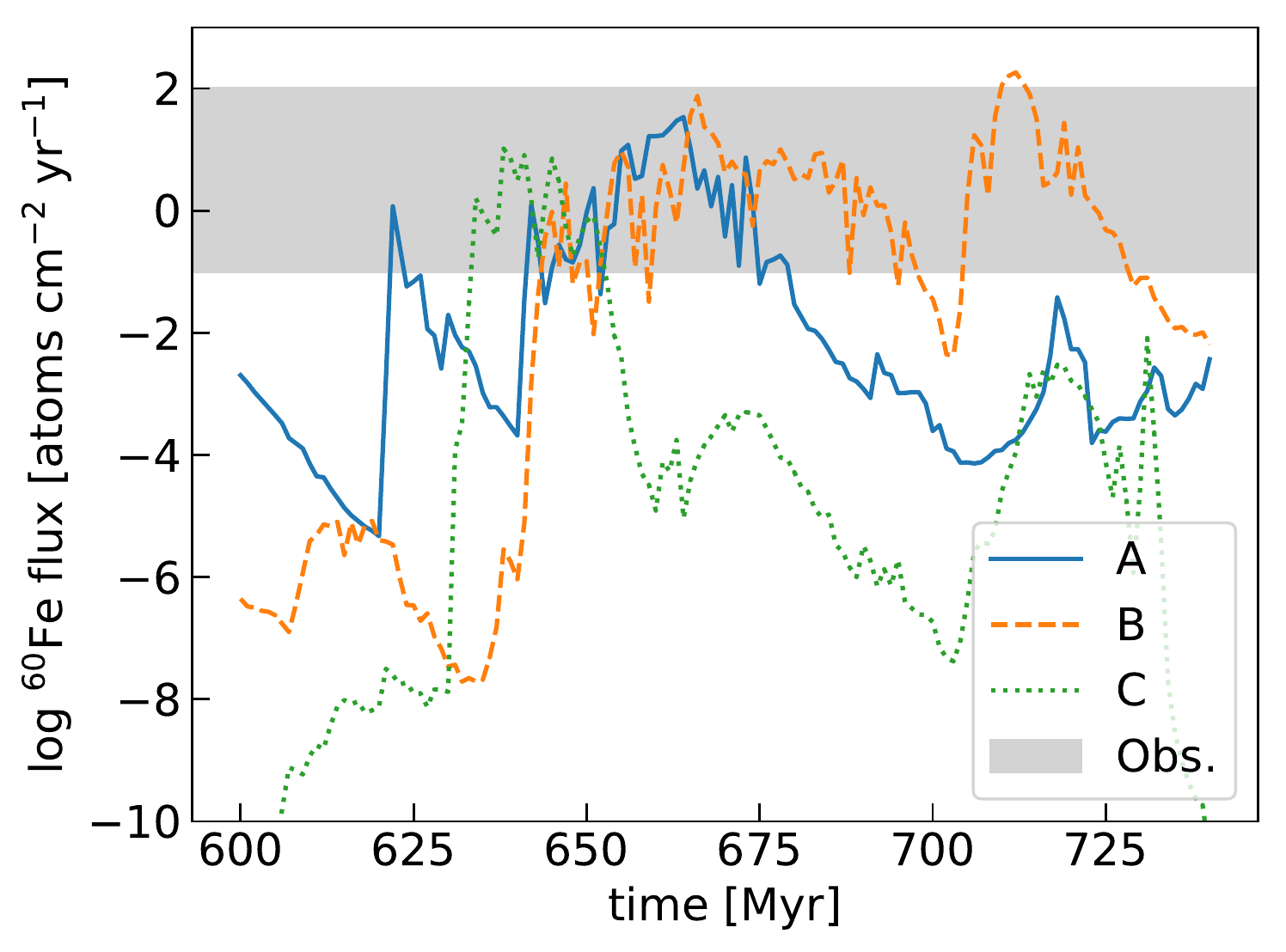}
	\caption{Time evolution of the $^{60}\text{Fe}$ flux that each of the stars A, B, and C shown in \autoref{fig: time_evolution_bubbles} receives. The observed range for the Earth is shown in grey.}
    \label{fig: 60Fe_influx_history}
\end{figure}

We now investigate the properties and evolutionary history of Galactic environments that simultaneously satisfy the three constraints we have considered in the previous sections. \autoref{fig: projections} shows a face-on view of the whole galactic disc and zoom-in images of distributions of $^{60}\text{Fe}$, $^{26}\text{Al}$, and X-ray emissivity, overlaid with the positions of stars that meet the three conditions. It clearly shows that stars that match the levels of SLRs and X-rays that we see from Earth are located exclusively inside kpc-scale bubbles that lie along the Galactic spiral arms, and are produced by massive stellar feedback. The bubble sizes are comparable to the widths of the gaseous spiral arms and one order of magnitude larger than the sizes of the individual giant molecular clouds (GMCs).
\autoref{fig: time_evolution_bubbles} shows the time evolution of the three stars and environments that we have identified as matching our observational constraints. In the figure, we show only the gas and $^{60}\text{Fe}$ because $^{26}\text{Al}$ and X-ray distributions are qualitatively similar to the $^{60}\text{Fe}$.

Examining \autoref{fig: time_evolution_bubbles}, we can identify a few common features in all three cases. First, the bubbles of gas in which the sample stars are located at 650 Myr (the time at which we select them) are relatively long-lived -- the bubbles present at 650 Myr are clearly identifiable for many tens of Myr before and after this point, so that the overall lifetime of the bubble is $\gtrsim 100$ Myr; this is long compared to the lifetime of any individual massive star, and is a result of a continuous supply of gas to fuel new star formation that is provided by the spiral arm. However, this does not mean that the sample star remains within the bubble for this entire time. In all cases the stars undergo epicyclic motion that is not identical to the motion of the gas that fuels the ongoing star formation. In case B, the gas and stellar motions are closely aligned, so that the star remains within the bubble for $\approx 90$ Myr, almost as long as the lifetime of the bubble itself. For C the duration of overlap is much shorter, with the star essentially plunging through the spiral arm and bubble, requiring only $\approx 20$ Myr to transit. Case A is intermediate. We illustrate the differences between these cases in \autoref{fig: 60Fe_influx_history}, which shows the time history of the $^{60}{\rm Fe}$ flux experienced by each star. Clearly a range of exposure durations from $\sim 10 - 100$ Myr is possible, depending primarily on the stellar orbit relative to the spiral arm.

\section{Conclusions and Discussion}
\label{Conclusions and Discussion}

Using an $N$-body+hydrodynamics simulation of a Milky-Way-like galaxy, we have investigated the location of stars on Sun-like orbits whose environments are consistent with three observational constraints seen from Earth: the $^{60}\text{Fe}$ influx onto the Earth's surface detected in deep-sea archives and Antarctic snow, a broad distribution of $^{26}\text{Al}$ observed in the $\gamma$-ray sky-maps, and the mean flux of diffuse soft X-ray emission. We find that stars that meet all three constraints are uncommon but not exceptionally rare; the number is 3 out of 151 total sample stars in our simulation, corresponding to a fraction of 2 per cent. Such stars are found predominantly inside kpc-scale bubbles of hot gas that are blown by feedback from massive stars, which form on the spiral arms.

We look into the time evolution of the three stars and investigate the formation and evolution of the local interstellar environment. 
We find that the time for which stars reside in feedback-blown bubbles is governed by the crossing time of stars across the spiral arm, which is $\sim 10 - 100$ Myr depending on the stellar trajectory. On the other hand, the residence time is insensitive to the lifetime of the bubble itself: in all the cases we identify where a bubble gives rise to an observed interstellar environment similar to that of the Sun, the bubble of hot gas that is responsible has a lifetime of $\approx 100$ Myr as a result of continuous fuelling of star formation by Galactic-scale spiral flows. Both the bubble lifetime and the residence time of a star within the bubble are much longer than the $\sim 10$ Myr lifetime of a single generation of massive stars; bubbles large enough to produce observational signatures similar to the ones we observe in the Solar System are the product of ongoing, multi-generational star formation, not a single cluster or burst.

In the Milky Way, the residence time of the Sun in the Local Arm (or Orion Arm) may be on the order of $\sim$ 60 Myr, considering the Solar velocity of 16 km s$^{-1}$ with respect to the Local Standard of Rest (LSR) and the $\sim 1$ kpc width of the Local Arm \citep{Bland-Hawthorn_Gerhard_2016}. This is uncertain because we do not know the velocity of the Local Arm, but it is in reasonable agreement with our simulation. Although the $\gamma$-ray and X-ray observations give us only instantaneous properties of the local ISM, $^{60}\text{Fe}$ in deep-sea archives and Antarctic snow should provide information about the Earth's history of exposure to $^{60}\text{Fe}$ over long timescales. Currently, at most 10 Myr of the history of $^{60}\text{Fe}$ influx has been investigated \citep[e.g.][]{Wallner_et_al_2016}, but studies probing at least 20 Myr are needed to confirm our results and to further understand the past and future of the local interstellar environment.

We also caution that our simulation may underestimate the residence times of stars in bubbles. In the simulation, the orbits of stars deviate from simple epicyclic motion that is expected in a smooth gravitational potential, and this deviation should be due to gravitational interactions with stars and gas clouds. As is always the case for $N$-body simulations of galaxies, our simulation contains far fewer stars than the actual number present in our Galaxy, and hence has a two-body relaxation timescale that is artificially small. As a result, our simulation overestimates the strength of gravitational scattering by the stellar components, which might artificially inflate the deviations of star particle orbits from simple epicycles. If this is the case, then in reality we expect to find slightly more stars that remain in the arms where conditions are favourable for accumulation of $^{60}\text{Fe}$ and $^{26}\text{Al}$. We do not expect this to be a large effect. However, more accurate description of the actual orbits of stars and long-term migration process requires technical improvement in our models of the stellar components of the disc, which will be the subject of future work.

\section*{Acknowledgements}

MRK acknowledges support from the Australian Research Council through \textit{Future Fellowship} FT180100375. SI is supported by JSPS KAKENHI Grant Numbers 16H02160, 18H05436, and 18H05437. Simulations were carried out on the Cray XC50 at the Center for Computational Astrophysics (CfCA) of the National Astronomical Observatory of Japan, the Gadi at the National Computational Infrastructure (NCI), which is supported by the Australian Government, and Oakforest-PACS provided by Multidisciplinary Cooperative Research Program in Center for Computational Sciences, University of Tsukuba. 

\section*{Data availability}
The data underlying this article will be shared on reasonable request to the corresponding author. 




\bibliographystyle{mnras}
\bibliography{reference} 




\appendix

\section{Dust-gas coupling}
\label{Dust-gas coupling}

We wish to estimate the characteristic speed with which dust grains of characteristic size $R_{\rm d}$ and internal density $\rho_{\rm d}$ will drift with respect to the gas in the diffuse interstellar medium, characterised by number density $\lesssim 10$ cm$^{-3}$ (roughly our star formation threshold in the simulation). Our discussion here follows the general treatment of the problem given by \citet{Hopkins_Squire_2018, Hopkins_Squire_2018b}.

The characteristic speed with which grains drift relative to the gas is given by
\begin{equation}
    v_{\rm drift} \approx a t_{\rm S},
\end{equation}
where $a$ is the acceleration (applied separately to the grains or the gas) that is responsible for causing the drift, and $t_{\rm S}$ is the stopping time that characterises the forces coupling dust and gas. In a galactic disc, the acceleration responsible for decoupling dust and gas arises from hydrodynamic forces that act on gas but not grains, and their characteristic amplitude must be of order the gas velocity divided by the characteristic timescale of the flow; this is $a\sim \sigma_{\rm g}/(h/\sigma_{\rm g}) = \sigma_{\rm g}^2/h$, where $\sigma_{\rm g}$ is the gas velocity dispersion and $h$ is the scale height of the neutral ISM. We can therefore write the ratio of the drift speed to the gas velocity dispersion, which is the quantity of interest for us, as
\begin{equation}
    \frac{v_{\rm drift}}{\sigma_{\rm g}} \approx \frac{\sigma_{\rm g} t_{\rm S}}{h} = 0.048 \left(\frac{\sigma_{\rm g}}{7\mbox{ km s}^{-1}}\right) \left(\frac{h}{150\mbox{ pc}}\right)^{-1} \left(\frac{t_{\rm S}}{\mbox{Myr}}\right),
\end{equation}
where the values of $\sigma_{\rm g}$ and $h$ to which we have scaled are the observed values in the Solar neighbourhood \citep{Boulares_Cox_1990}. Grain-gas drift is dynamically significant only if this ratio approaches unity, and our approximation of neglecting it is reasonable as long as the stopping time $t_{\rm S} \lesssim 10$ Myr.

Under diffuse ISM conditions, we must consider a range of possible coupling processes: collisional drag, Coulomb drag, and Lorentz forces. We can estimate each of these in turn. Collisional drag is strongly in the Epstein regime, since the particle mean free path is of order AU, and the stopping time is of order
\begin{equation}
    t_{\rm Eps} = \left(\frac{\pi \gamma}{8}\right)^{1/2} \frac{\rho_{\rm d} R_{\rm d}}{n_{\rm H} \mu m_{\rm H} c_{\rm s}} = \left(\frac{\pi \gamma}{8}\right)^{1/2} \frac{\rho_{\rm d} R_{\rm d} c_{\rm s}}{P},
\end{equation}
where $\mu_{\rm H}=1.4$ is the mean mass per H nucleus in units of the hydrogen mass $m_{\rm H}$, and $\gamma$, $P$, and $c_{\rm s}$ are the adiabatic index, thermal pressure, and sound speed of the ISM. The radiative cooling time of the ISM is generally short compared to mechanical timescales, so we can set $\gamma\approx 1$. We therefore have
\begin{align}
    t_{\rm Eps} &\approx 1.4\mbox{ Myr}
    \nonumber \\
    &\times \left(\frac{\rho_{\rm d}}{3\mbox{ g cm}^{-3}}\right) \left(\frac{R_{\rm d}}{1\;\mu\mathrm{m}}\right) \left(\frac{c_{\rm s}}{1\mbox{ km s}^{-1}}\right)\left(\frac{P/k_{\rm B}}{3000\mbox{ K cm}^{-3}}\right)^{-1}.
    \label{eq:t_eps}
\end{align}
The value of $\rho_{\rm d}$ to which we have scaled here is appropriate for rocky materials, the pressure (normalised by Boltzmann's constant $k_{\rm B}$) is characteristic of the Milky Way's diffuse ISM \citep{Wolfire_2003}, and the sound speed to which we have scaled is appropriate for the cold phase of the diffuse ISM; for the warm phase, $c_{\rm s}$ would be a factor of $4-5$ larger.

Coulomb and Lorentz forces depend on the grain charge. Grains in the size range with which we are concerned ($\sim 0.01 - 10$ $\mu$m) in the diffuse ISM tend to be positively charged as a result of photoelectric ejection. The charge state is a set by the balance between this process and recombination with free electrons, and over the size range with which we are concerned can be approximated by \citep{Tielens_2005}
\begin{align}
    Z_{\rm d} &\approx -1 + (f_{\rm L} - 1) \left(\frac{R_{\rm d} k_{\rm B} T}{e^2}\right) \\
    &\approx  60\ (f_{\rm L} - 1) \left(\frac{R_{\rm d}}{1\ \mu \rm m}\right) \left(\frac{T}{1000\ \rm K}\right),
\end{align}
where $e$ is the elementary charge, $Z_{\rm d}$ is the grain charge in units of $e$, $f_{\rm L}$ is a number that characterises the local FUV radiation field strength and electron density, and is typically of order few for diffuse ISM conditions, $T$ is the gas temperature, and in the numerical evaluation we have dropped the leading $-1$ since it is generally unimportant. Coulomb drag, assuming the dominant ions with which the grains are interacting are protons (appropriate for the diffuse ISM) then gives a stopping time
\begin{equation}
    t_{\rm C} \approx \left(\frac{\pi \gamma}{2}\right)^{1/2} \frac{\rho_{\rm d} R_{\rm d} c_{\rm s}}{f_{\rm ion} P \ln\Lambda} \left(\frac{R_{\rm d} k_B T}{e^2 Z_{\rm d}}\right)^2 \approx \frac{2}{(f_{\rm L}-1)^2 f_{\rm ion} \ln\Lambda} t_{\rm Eps},
    \label{eq:t_c}
\end{equation}
where $f_{\rm ion}$ is the ionisation fraction and $\ln\Lambda$ is a Coulomb logarithm. Ionisation fractions in the diffuse ISM take on values in the range $f_{\rm ion} \sim 10^{-3} - 10^{-2}$ \citep{Wolfire_2003}, and $\ln \Lambda \sim 15-20$ under astrophysical conditions \citep{Hopkins_Squire_2018b}, so the pre-factor in front ot $t_{\rm Eps}$ in \autoref{eq:t_c} is generally of order $1-10$; we can therefore consider Coulomb drag to be comparable to or weaker than Epstein drag in importance.

Finally, the stopping time for magnetic forces is the Larmor time,
\begin{align}
    t_{\rm L} & = \frac{4\pi \rho_{\rm d} R_{\rm d}^3 c}{3 e Z_{\rm d} B} \approx \sqrt{\frac{2\pi\beta}{9}} \frac{c e\rho_{\rm d} R_{\rm d}^2}{(f_{\rm L}-1) \mu m_{\rm H} c_{\rm s}^2 \sqrt{P}} \\
    & = 0.8\mbox{ Myr} \; \frac{\sqrt{\beta}}{f_{\rm L}-1} \left(\frac{\rho_{\rm d}}{3\mbox{ g cm}^{-3}}\right) \left(\frac{R_{\rm d}}{1\;\mu\mathrm{m}}\right)^2
    \nonumber \\
    &\qquad \times \left(\frac{c_{\rm s}}{1\mbox{ km s}^{-1}}\right)^{-2}  \left(\frac{P/k_{\rm B}}{3000\mbox{ K cm}^{-3}}\right)^{-1/2},
    \label{eq:t_L}
\end{align}
where $\mu = 1.3$ is the mean mass per free particle in units of the hydrogen mass, $B$ is the magnetic field, and $\beta = 8\pi P/B^2$ is the plasma $\beta$, typically $\sim 1$ in the diffuse ISM \citep{Boulares_Cox_1990}.

The stopping time will generally be the minimum of the three timescales $t_{\rm Eps}$, $t_{\rm C}$, and $t_{\rm L}$ that we have computed. Examining \autoref{eq:t_eps} and \autoref{eq:t_L}, we see that, for the cold phase of the neutral ISM ($c_{\rm s} \approx 1$ km s$^{-1}$), we expect Epstein drag to dominate for grains of size $R_{\rm d} \gtrsim 2-3$ $\mu$m (depending on the numerical value adopted for $f_{\rm L}-1$), and that our condition $t_{\rm S} \lesssim 10$ Myr is then satisfied for grains up to $R_{\rm d}\approx 10$ $\mu$m in size. For the warm phase ($c_{\rm s} \approx 5$ km s$^{-1}$), Lorentz coupling dominates, and produces $t_{\rm S} \lesssim 10$ Myr for grain sizes $R_{\rm d} \lesssim 20-30$ $\mu$m, again depending on the exact numerical value of $f_{\rm L}-1$. Thus we generically expect that our neglect of grain drift with respect to gas is reasonable for grains up to $\sim 10$ $\mu$m in size.


\bsp	
\label{lastpage}
\end{document}